\documentclass[aip,graphicx,reprint,amsmath,amssymb]{revtex4-2}

\usepackage{graphicx}
\DeclareUnicodeCharacter{2009}{\,} 
\DeclareUnicodeCharacter{2212}{--}
\DeclareUnicodeCharacter{2A7D}{$\leq$}
\DeclareUnicodeCharacter{03B2}{$\beta$}
\usepackage{dcolumn}
\usepackage{bm}
\usepackage[mathlines]{lineno}

\usepackage[utf8]{inputenc}
\usepackage[T1]{fontenc}
\usepackage{mathptmx}
\usepackage{etoolbox}

\begin{document}
	
	
	\title[Advances in high-pressure laser floating zone growth: the Laser Optical Kristallmacher II (LOKII)]{Advances in high-pressure laser floating zone growth: the Laser Optical Kristallmacher II (LOKII)} 
	
	
	\author{Steven J. Gomez Alvarado}
	\altaffiliation{stevenjgomez@ucsb.edu}
	\affiliation{Materials Department, University of California, Santa Barbara, CA 93106-5050, USA}
	\author{Eli Zoghlin}
	\affiliation{Materials Department, University of California, Santa Barbara, CA 93106-5050, USA}
	\affiliation{William H. Miller III Department of Physics and Astronomy, Johns Hopkins University, Baltimore, Maryland 21218, USA}
	\author{Azzedin Jackson}
	\author{Linus Kautzsch}
	\affiliation{Materials Department, University of California, Santa Barbara, CA 93106-5050, USA}
	\author{Jayden Plumb}
	\affiliation{Materials Department, University of California, Santa Barbara, CA 93106-5050, USA}
	\affiliation{Department of Mechanical Engineering, University of California, Santa Barbara, CA 93106-5050, USA}
	\author{Michael Aling}
	\affiliation{Materials Department, University of California, Santa Barbara, CA 93106-5050, USA}
	\affiliation{Department of Mechanical Engineering, Massachusetts Institute of Technology, Cambridge, MA 02139, USA}
	\author{Andrea N. Capa Salinas}
	\author{Ganesh Pokharel}
	\author{Yiming Pang}
	\author{Reina M. Gomez}
	\affiliation{Materials Department, University of California, Santa Barbara, CA 93106-5050, USA}
	\author{Samantha Daly}
	\affiliation{Department of Mechanical Engineering, University of California, Santa Barbara, CA 93106-5050, USA}
	\author{Stephen D. Wilson}
	\altaffiliation{stephendwilson@ucsb.edu}
	\affiliation{Materials Department, University of California, Santa Barbara, CA 93106-5050, USA}
	
	
	
	\date{\today}
	
	\begin{abstract}
		The optical floating zone crystal growth technique is a well-established method for obtaining large, high-purity single crystals. While the floating zone method has been constantly evolving for over six decades, the development of high-pressure (up to 1000~bar) growth systems has only recently been realized via the combination of laser-based heating sources with an all-metal chamber. While our inaugural high-pressure laser floating zone furnace design demonstrated the successful growth of new volatile and metastable phases, the furnace design faces several limitations with imaging quality, heating profile control, and chamber cooling power. 
		Here, we present a second-generation design of the high-pressure laser floating zone furnace, ``Laser Optical Kristallmacher II'' (LOKII), and demonstrate that this redesign facilitates new advances in crystal growth by highlighting several exemplar materials: $\alpha$-Fe$_2$O$_3$, $\beta$-Ga$_2$O$_3$, and La$_2$CuO$_{4+\delta}$. Notably, for La$_2$CuO$_{4+\delta}$, we demonstrate the feasibility and long-term stability of traveling solvent floating zone growth under a record pressure of 700~bar.

	\end{abstract}
	
	
	\maketitle 
	
	
	\section{Introduction}
	
	For over 60 years, the optical floating zone crystal growth method \cite{de_la_rue_arcimage_1960} has played a critical role in the development of pristine single crystals for the study of novel electronic, optical, and magnetic phenomena. Despite being a well-established technique, floating zone growth remains an ever-developing frontier among active crystal growth methods.\cite{schmehr_active_2017} Among recent advances in optical floating zone (OFZ) technology, the application of elevated gas pressures during growth (up to 300~bar in commercially available systems produced by ScIDre, GmbH) has been particularly effective in expanding the phase space of compounds which can be grown via the floating zone method.\cite{balbashov_apparatus_1981,souptel_vertical_2007} 
	Since the inception of the high-pressure laser floating zone (HP-LFZ) furnace \cite{schmehr_high-pressure_2019}, which allows for the application of pressures up to 1000~bar using an all-metal growth chamber, the HP-LFZ has proven effective in suppressing material volatility and modifying chemical potential during growth.\cite{zoghlin_evaluating_2021,porter_crystal_2022,zoghlin_refined_2023}
	Our inaugural design of the HP-LFZ furnace has presented challenges with imaging quality, sample heating, and chamber cooling power --- issues that are amplified at the highest operating pressures.
	Here, we present a second-generation of the HP-LFZ furnace based on a redesign of the chamber geometry which allows for in-plane viewing during growth and improved cooling of the chamber.  These design modifications are embodied in a modified furnace named Laser Optical Kristallmacher II (LOKII). We detail the changes in this furnace design and demonstrate that these improvements facilitate the exploration of new regimes of OFZ crystal growth using several materials as examples.

	\section{Instrument description}
	
	\subsection{Chamber design}
	
	\begin{figure*}[htb!]
		\centering
		\includegraphics[width=0.9\linewidth]{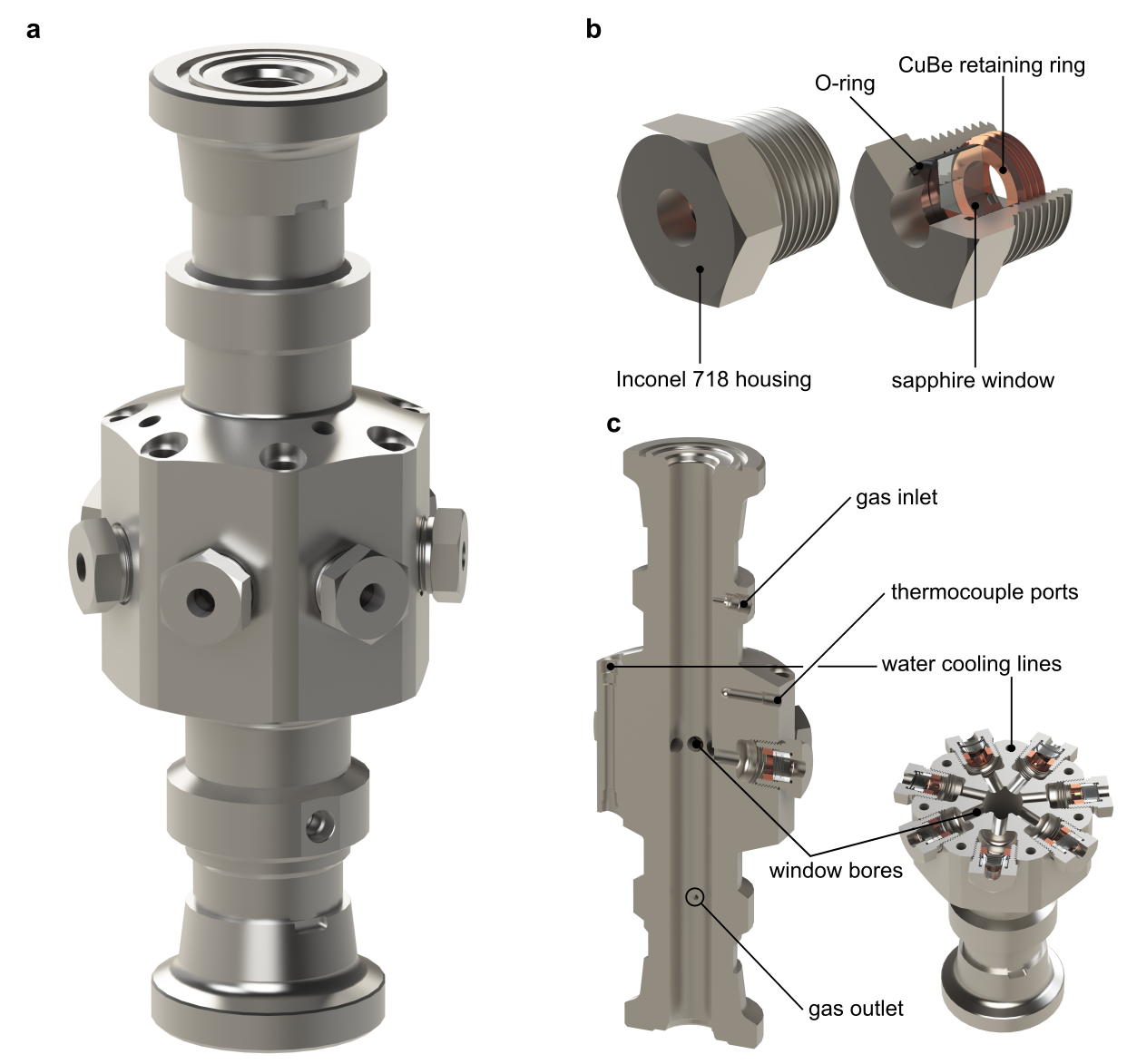}
		\caption{\label{fig:SightGlassAndChamber}(a) Rendering of the second-generation chamber design. (b) Rendering of the sight glass housings with a cut view highlighting the O-ring and retaining ring which create a seal about the sapphire window. (c) Cutaways of the chamber rendering revealing additional water cooling lines for improved cooling power and enlarged window bores for enhanced optical access.}
	\end{figure*}
	
\begin{figure*}[htb!]
		\centering
		\includegraphics[width=0.85\linewidth]{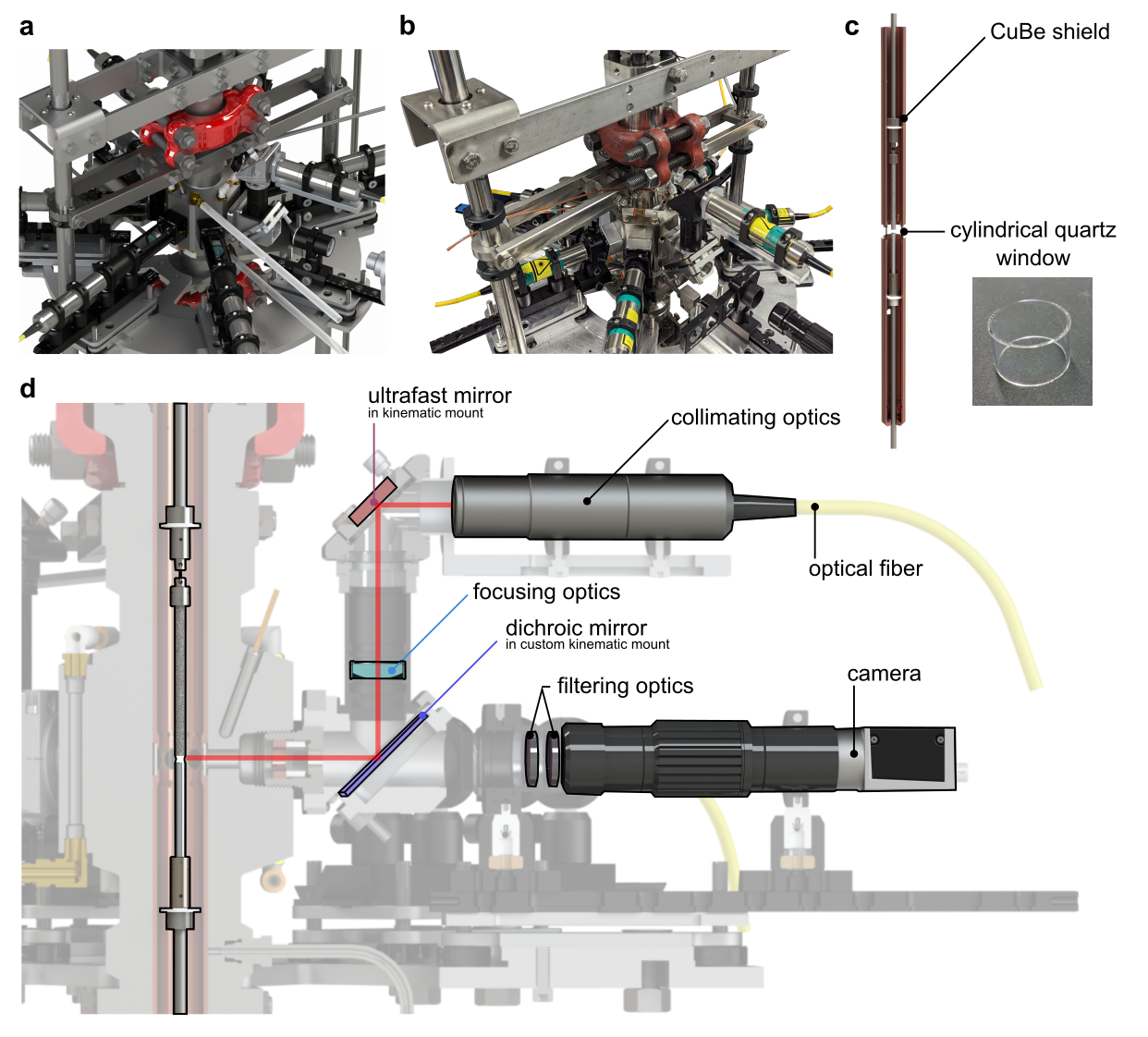}
		\caption{\label{fig:IPV}(a) Rendering of the full HP-LFZ apparatus with all components excluding the external frame. (b) A photograph of the HP-LFZ apparatus for comparison. (c) Rendering of the second-generation heat and chemical shield. A cylindrical window between the CuBe shield pieces allows for laser transmission and imaging. (d) A cut view of the in-plane viewing optics array with key components highlighted.}
	\end{figure*}

	The new chamber design (Fig.~\ref{fig:SightGlassAndChamber}) features several key differences from the inaugural design of the HP-LFZ. Firstly, the original 45$^{\circ}$ out-of-plane sight glasses used for imaging and pyrometry were removed. Viewing of the zone is accomplished by making use of semi-transparent, laser-reflective optics that allow for imaging and pyrometry directly down the line of sight of the lasers (Fig.~\ref{fig:IPV}). This in-plane viewing geometry dramatically improves imaging quality for volatile growths at the highest pressures and temperatures, where the rising turbulent flow of gas inside the chamber would inhibit visual monitoring of the growth in the out-of-plane viewing geometry. With the removal of the out-of-plane bores, the in-plane bores now feature an increased diameter of 9~mm allowing for a wider field of view for imaging.
	
	Secondly, this chamber features a custom sight glass design which is composed of a window housing made of UNS~N07718 (Inconel 718) and an anti-reflection (AR) coated sapphire sight glass (Guild Optical Associates, Inc.). The pressure seal is achieved by a retaining ring made of CuBe alloy UNS~C17200 which holds the sapphire window against a Viton O-ring (90A Duro FKM, Marco Rubber). This window design is capable of withstanding the highest operating pressures, though the O-ring is susceptible to softening and extrusion upon overheating of the chamber (above 100 $^{\circ}$C), which causes a failure of the pressure seal. 
	
	To remedy this issue, we have included additional cooling water lines in the new chamber, paired with a 1000~W water-to-water chiller (LX1 Series, Haskris). The increased cooling power is especially important at higher pressures, where increased fluid density enhances the heat transfer away from the sample. The removable heat and chemical shield has also been redesigned (Fig.~\ref{fig:IPV}(c)), featuring a single cylindrical quartz window which provides a wide field of view for imaging and transmission of laser light to the sample. Previously made of aluminum, the shield is now also made of the CuBe alloy UNS~C17200 and is capable of resisting chemical attack even at the highest operating temperatures and O$_2$ pressures.  An 80:20 Ar:O$_2$ gas composition is the typical oxygen source used for high-pressure growths in the furnace for administrative safety reasons, as this partially mitigates the hazards of unknown contamination within an initially oxygen clean gas service and in maintaining a booster capable of pushing pure O$_2$ above 5000 psi.

	\begin{figure}[htb!]
		\centering
		\includegraphics[width=0.85\linewidth]{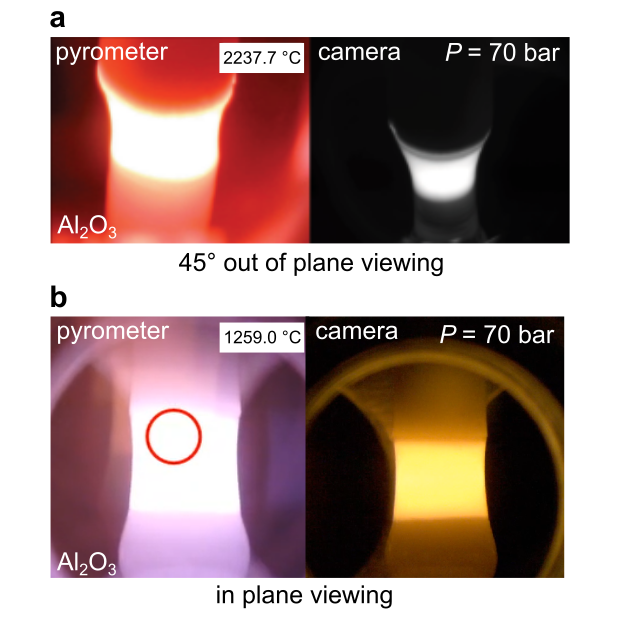}
		\caption{\label{fig:Imaging}(a) Images from a growth of Al$_2$O$_3$ under $P=70$~bar from the pyrometer (left) and camera (right), viewed through the out-of-plane sight glasses in the old chamber design (Adapted from Ref. 5). (b) Corresponding images of a growth of Al$_2$O$_3$ repeated in the new chamber design, viewed through in-plane sight glasses. A systematic difference in the measured temperature is due to filtering effects caused by a laser line mirror which sits in the line of sight of the pyrometer.}
	\end{figure}

	\subsection{Laser Heating Source}
	
	In our inaugural furnace design, the HP-LFZ relies on seven diode lasers, each capable of providing 100~W of infrared laser light at a wavelength of 820~nm. After attempting growths of a wide array of materials, 100~W per laser (700~W total) was found to be insufficient to melt highly refractory and thermally conducting compounds, particularly for materials with poor optical absorption at 820~nm. In addition, the design requirement of coupling numerous diode outputs into a collective fiber bundle creates a larger optical source with a correspondingly larger axial divergence.  The resulting inability to tightly collimate the beam created a lower bound on the height of the incident laser spot at the sample position of 2.1~mm in our previous design, limiting control over the sharpness of the heating profile.\cite{schmehr_high-pressure_2019} Finally, when growing under the highest pressures with significant material volatility, the increased turbulence causes fluctuations in the density of the fluid which, in turn, cause fluctuations in the index of refraction. The gradient of the index of refraction results in a redirection and broadening of the incident laser light, which decreases the amount of power transferred to the sample. 
	
	To circumvent these limitations, we constructed a second furnace (LOKII) using seven fiber lasers that are each capable of emitting 200~W of laser light at a wavelength of 1070~nm (IPG Photonics). Beyond the enhanced power output, these lasers also feature a negligible divergence of the beam after collimation, resulting in an unmodified spot size of 5~mm diameter at the sample position. Using a conventional cylindrical focusing lens, the height of this spot can be focused down to a line, enabling enhanced control over the heating profile. Additionally, with a reduced incident spot size (i.e., increased power density), higher melting points can be accessed at lower laser powers. Together with the doubling of the maximum power output, these improvements expand the available range of accessible melting temperatures, especially pertinent for high-melting compounds such as refractory oxides, nitrides, and borides. We find that the issue of spot broadening and redirection via refraction during growths of volatile materials can be significantly diminished by narrowing the spot size below $\approx$1.5~mm, which we attribute to a minimization of the overlap between the turbulent layer of gas flow and the incident laser beam.

	\subsection{Imaging and Pyrometry}
	
	To grant the camera and pyrometer optical access via the same windows through which the laser light enters the chamber, we employ optics which allow for in-plane imaging of the molten zone. While five lasers remain in the plane, two lasers on opposite sides of the chamber are reflected into the chamber by semi-transparent, laser-reflective optics allowing for transmission of visible light into a camera for imaging and into a pyrometer for both imaging and temperature readings (Fig.~\ref{fig:IPV}(d)). An alternative design where laser light is transmitted and the optical image is reflected was rejected, as transmission of the laser light would have resulted in more losses than reflection of the laser light. The 820~nm furnace utilizes a Quantalux CS2100M-USB sCMOS camera from Thorlabs, Inc. and the 1070~nm furnace utilizes an acA1300-75gcE Color GigE Camera from Basler, both employing an Infiniprobe MS variable magnification lens and $2\times$ magnification lens from Infinity Photo-Optical Company. Both furnaces employ a Fluke Endurance pyrometer, model E1RH-F0-V-0-0.
	
	In this scheme, shown in Fig.~\ref{fig:IPV}(d), the laser light first leaves the laser fiber bundle and is focused by a collimating lens (IPG Photonics). It is then reflected 90$^{\circ}$ by a mirror in a kinematic mount (Thorlabs), followed by complementary 90$^{\circ}$ reflection off of a semi-transparent mirror, which allows for transmission of visible light but reflects at the laser wavelength. This mirror is mounted in a custom kinematic mount which minimizes the path length extension for the laser, thus reducing differences in the incident laser spot size between the modified and unmodified laser geometries. For the camera, this semi-transparent mirror is a dichroic mirror (Thorlabs), which reflects wavelengths larger than 750~nm. For the pyrometer, this semi-transparent mirror is a laser line mirror which reflects light at a range of wavelengths tightly centered around either 820~nm (CVI Laser Optics, LLC) or 1070~nm (Edmund Optics), respectively for each of the two furnaces. The pyrometer measurement relies on reading wavelengths in the infrared regime which include these bands, thus causing the measured temperature to be systematically off by several hundreds of degrees.
	
	Example images of the molten zone in Fig.~\ref{fig:Imaging} show a comparison between the previous out-of-plane and the new in-plane viewing geometries. At higher pressures, the imaging is significantly improved, which we attribute to a shallower viewing angle where there is a significantly smaller volume of turbulent fluid between the sight glass and the molten zone. 
	
	Previously, compounds with significantly volatile components (e.g., RuO$_2$, RhO$_2$, and IrO$_2$) would deposit a thick layer of powder on the viewing windows after just a few hours of growth, precluding visual monitoring of the zone. Now, continuous visual monitoring of the molten zone is made easier as laser heating prevents volatilized material from cooling down enough to deposit on the cylindrical window in the regions where the visual monitoring is being conducted.

	\section{Crystal Growth Demonstrations}
	
	\subsection{Tuning Phase Stability in $\alpha$-Fe$_2$O$_3$}
	
	\begin{figure*}[htb!]
		\centering
		\includegraphics[width=0.85\linewidth]{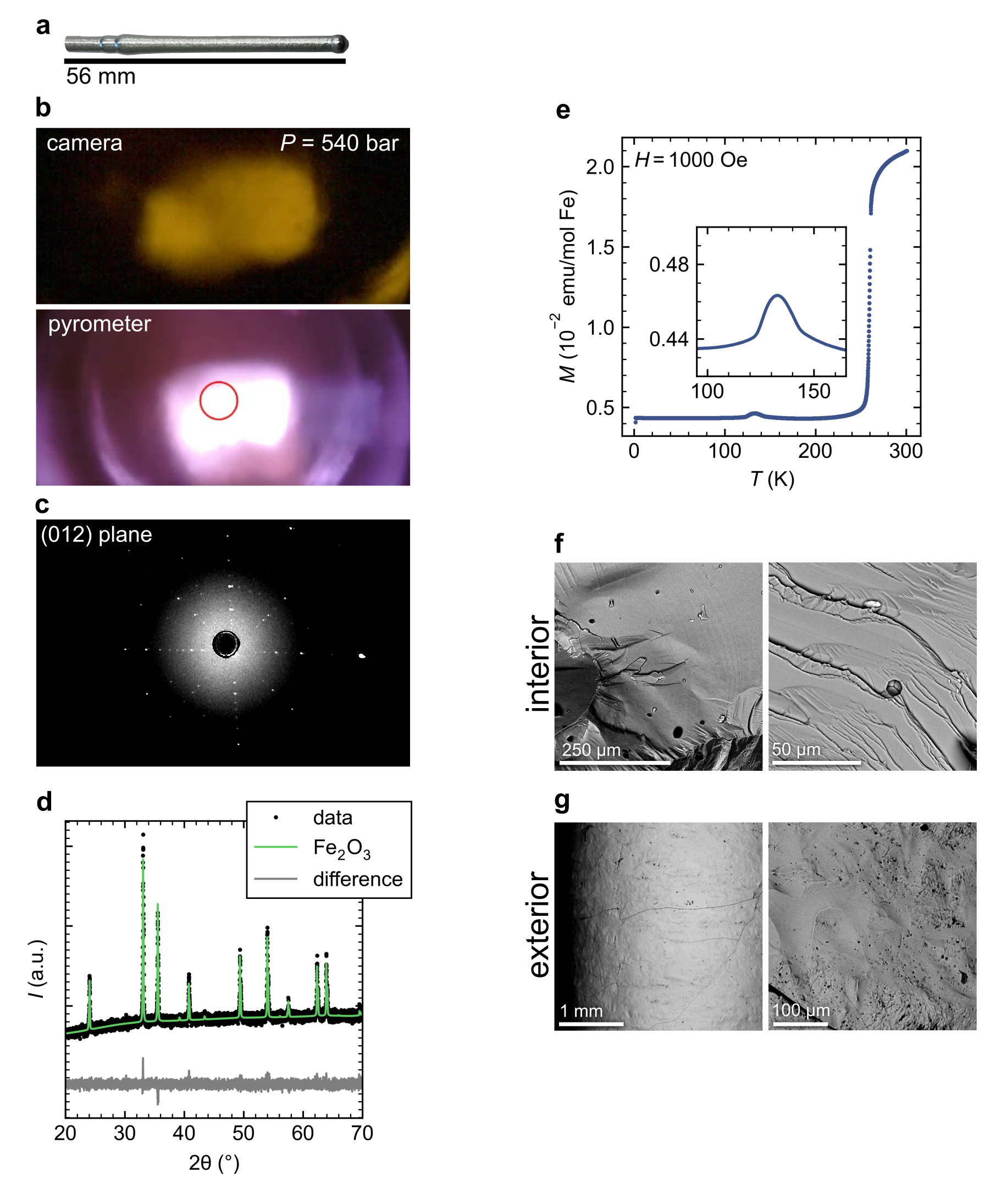}
		\caption{\label{fig:Fe2O3}(a) Single crystal of $\alpha$-Fe$_2$O$_3$ obtained from the HP-LFZ --- a shiny grey boule with no visible facets. (b) Images from the pyrometer and camera during the growth. (c) Laue diffraction pattern obtained from the boule, consistent with reflections from the (012) plane. (d) Powder X-ray diffraction data on a crushed portion of the crystal. All peaks can be indexed using the rhombohedral structure of $\alpha$-Fe$_2$O$_3$. (e) Temperature-dependent field-cooled magnetization data collected under a field of $H=1000$~Oe on a portion of the same powder used for X-ray diffraction. Inset shows the impurity feature corresponding to an estimated 0.085~mol\% of Fe$_3$O$_4$. (f) Back-scattered scanning electron micrographs of the interior of the boule, obtained from mechanical cleavage of the grown crystal, and (g) of the exterior face of the boule.}
	\end{figure*}
	
	As a demonstration of the ability of the LOKII design to grow pressure-stabilized oxide phases, $\alpha$-Fe$_2$O$_3$ (hematite) crystals were grown.
	In recent years, antiferromagnets have become of interest in spintronics due to their inherent potential to achieve faster dynamics in data recording and manipulation when compared to ferromagnets.\cite{zhang_control_2022,jungwirth_antiferromagnetic_2016,xiong_antiferromagnetic_2022}  Specifically, antiferromagnets with easy-plane anisotropy, like $\alpha$-Fe$_2$O$_3$, provide an opportunity to study spin-orbit torque switching. While most of the studies of this material have been performed on thin films, bulk single crystals of $\alpha$-Fe$_2$O$_3$ are ideal due to their higher purity and well-defined orientation (i.e., single grain nature). 
	
	The growth of $\alpha$-Fe$_2$O$_3$ bulk single crystals at low pressure is made challenging by the difficulty of obtaining the correct oxidation state of Fe$^{3+}$, which results in incongruent melting. To date, most bulk single crystal samples of $\alpha$-Fe$_2$O$_3$ have been synthesized at low pressure using fluxes, including crucible-based techniques\cite{tasaki_magnetic_1963,chen_magnetic-field-induced_2012} and the traveling solvent floating zone method.\cite{chiaramonti_optical_2004} Small amounts of flux incorporation modify the magnetic properties of $\alpha$-Fe$_2$O$_3$, which are known to be sensitive to impurities. One of the key advantages of the HP-LFZ is the ability to routinely apply large overpressures of O$_2$ and thereby favorably modify the thermodynamics of the recrystallization process. Growth of $\alpha$-Fe$_2$O$_3$ provides an excellent example of this capability, as $\alpha$-Fe$_2$O$_3$ has been shown to melt congruently only at O$_2$ pressures above 50~bar.\cite{crouch1971magnetite}  High-pressure growth thus enables standard floating zone growth without need for a flux.\cite{balbashov_apparatus_1981}  
	
	High purity $\alpha$-Fe$_2$O$_3$ powder (Alfa Aesar, 99.995\%) was dried for $\approx$12 hours at 600~$^\circ$C and then packed inside cylindrical rubber balloons (diameter~$\approx$~6~mm) to make a seed rod (length~$\approx$~25~mm) and a feed rod (length~$\approx$~100~mm). These rods were pressed using a cold isostatic press at 3250~bar and subsequently sintered for 24~hours under O$_2$ flow at 1200~$^\circ$C. The growth parameters --- particularly gas pressure and laser power --- were initially optimized through several iterations using polycrystalline seed rods. 
	For the final growth using the optimized parameters, a single crystal from a previous growth was used as the seed. The growth was then conducted under 550~bar of an 80:20 Ar:O$_2$ atmosphere in order to stabilize congruent melting. Stable melting and recrystallization of the feed rod onto the seed was accomplished using a total applied power of 455~W, at which point the molten zone remained stable throughout the growth. Both feed and seed were translated at a rate of 8.5~mm/hr and counter-rotated at rates of 8~rpm and 9~rpm, respectively. No significant volatility of hematite was observed during growth.
	
	The resulting $\alpha$-Fe$_2$O$_3$ sample is shown in Fig.~\ref{fig:Fe2O3}(a). Laue X-ray diffraction (Fig.~\ref{fig:Fe2O3}(c)) revealed the successful formation of a single crystal with the growth direction approximately normal to the ($0\bar{1}\bar{2}$) plane. Powder X-ray diffraction on a crushed portion of the single crystal using a PANalytical Empyrean diffractometer suggests phase-pure $\alpha$-Fe$_2$O$_3$, (Fig. ~\ref{fig:Fe2O3}(d)). 
	
	However, measurement of the magnetic properties of a ground piece of the boule using a Quantum Design MPMS3 SQUID magnetometer revealed the presence of a small amount of ferromagnetic Fe$_3$O$_4$ (magnetite). The presence of Fe$_3$O$_4$ is indicated by the small feature near 133~K in the magnetization data (Fig.~\ref{fig:Fe2O3}(e)). This feature corresponds to the Verwey transition\cite{walz_verwey_2002}, which appears in single-phase Fe$_3$O$_4$ at $T_V\approx125$~K. The fact that we were not able to detect any of the Fe$_3$O$_4$ phase via X-ray diffraction sets an approximate upper bound of 1\% by mass in the sample, which is the estimated resolution limit of the powder diffractometer. To better quantify the amount of Fe$_3$O$_4$ in the sample, the height of the impurity peak from the magnetization versus temperature data was scaled to Fe$_3$O$_4$ magnetization data from literature.\cite{tasaki_magnetic_1963} This analysis suggests that the Fe$_3$O$_4$ content in the sample is approximately 0.085 mol\%, which is comparable to the amount observed in magnetization measurements performed on the starting material.
	
	\subsection{Growth of Wide Band Gap $\beta$-Ga$_2$O$_3$}
	
	\begin{figure}[t!]
		\centering
		\includegraphics[width=0.85\linewidth]{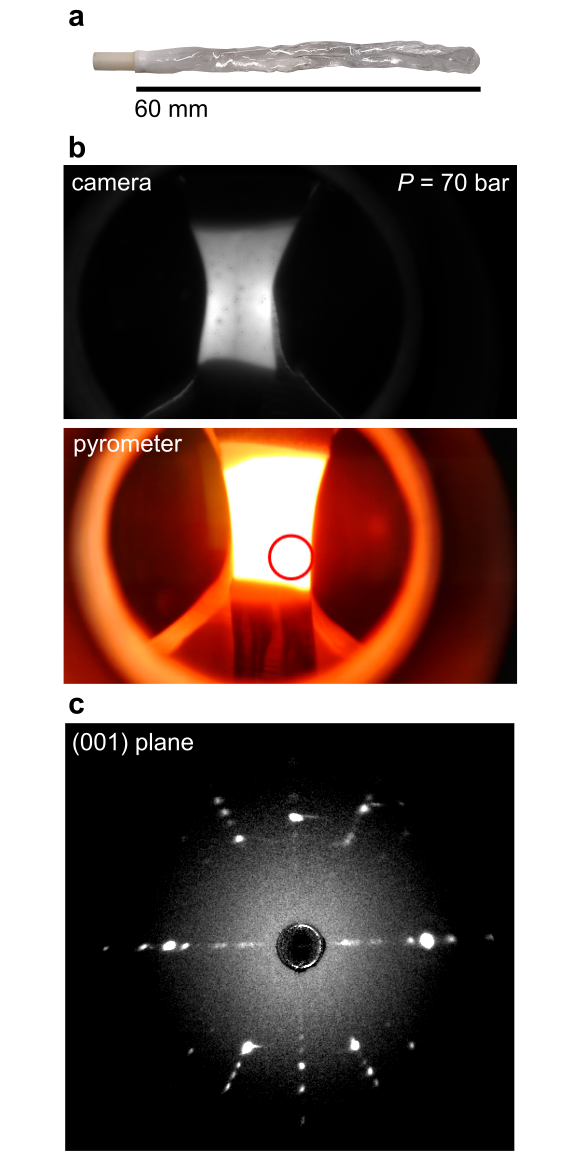}
		\caption{\label{fig:Ga2O3}(a) Single crystal of $\beta$-Ga$_2$O$_3$ obtained from the HP-LFZ --- a nearly transparent boule with noticeable facets. (b) Images from the pyrometer and camera during growth at $P=70$~bar. (c) Laue diffraction pattern obtained from the boule, consistent with reflections from the (001) plane.}
	\end{figure}

	As a demonstration of the ability for the relatively low frequency, monochromatic laser light of the LOKII design to couple to and melt large band gap insulators, growth of $\beta$-Ga$_2$O$_3$ was performed. $\beta$-Ga$_2$O$_3$ is a wide band gap (4.8~eV) material of interest for its applications in optoelectronics and power electronics. This compound has been grown using both conventional \cite{tomm_floating_2001, villora_large-size_2004} and CO$_2$ laser-based floating zone techniques.\cite{santos_optical_2012} Due to its wide band gap, $\beta$-Ga$_2$O$_3$ exhibits low absorptivity in the infrared, resulting in poor coupling to the 820~nm laser light used for this growth. This was demonstrated by heating a seed rod using up to 100~W of 820~nm light per laser (700~W total), which proved incapable of melting the sample. To circumvent this issue, a rod of Al$_2$O$_3$ (96.0~$\%$~-~99.8~$\%$, McMaster-Carr) was used as the seed for the growth, to serve as an initial susceptor of the laser light, and a rod of $\beta$-Ga$_2$O$_3$ was used as the feed. Starting powder of $\beta$-Ga$_2$O$_3$ (99.99 $\%$, Alfa Aesar) was dried overnight at 1200~$^\circ$C and then pressed into 4~mm diameter rods at 3275~bar using a cold isostatic press. The rods were then sintered at 1200~$^\circ$C for 10~hours in air.
	
	The Al$_2$O$_3$ was first melted using a total applied power of 216~W at 820~nm, and a feed of $\beta$-Ga$_2$O$_3$ was joined into the melt. Subsequent translation rapidly diluted any Al content out of the molten zone and resulted in a colorless crystal of $\beta$-Ga$_2$O$_3$ (Fig. \ref{fig:Ga2O3}(a)). The growth was carried out with a translation rate of 15~mm/hr and counter-rotation rate of 10~rpm for both the feed and seed rods, using 70~bar of an 80:20 Ar:O$_2$ atmosphere to mitigate volatility. The level of volatility observed at the moderate pressure of 70~bar was not found to be prohibitive for the growth.
	
	To assess the phase purity and crystal structure, a micron-scale piece of the resulting crystal was broken off and measured on a Kappa Apex II single-crystal diffractometer with a fast charge-coupled device (CCD) detector and a Mo source. The single crystal X-ray diffraction pattern was refined using the SHELX software package.\cite{sheldrick_it_2015}
	The refined crystal structure consists of two and three different crystallographic sites for Ga and O, respectively, and is consistent with the previously reported structure of $\beta$-Ga$_2$O$_3$.\cite{villora_large-size_2004,geller_crystal_1960}

	\subsection{High-Pressure Traveling Solvent Growth of La$_2$CuO$_{4+\delta}$}
	
	\begin{figure}[t!]
		\centering
		\includegraphics[width=0.80\linewidth]{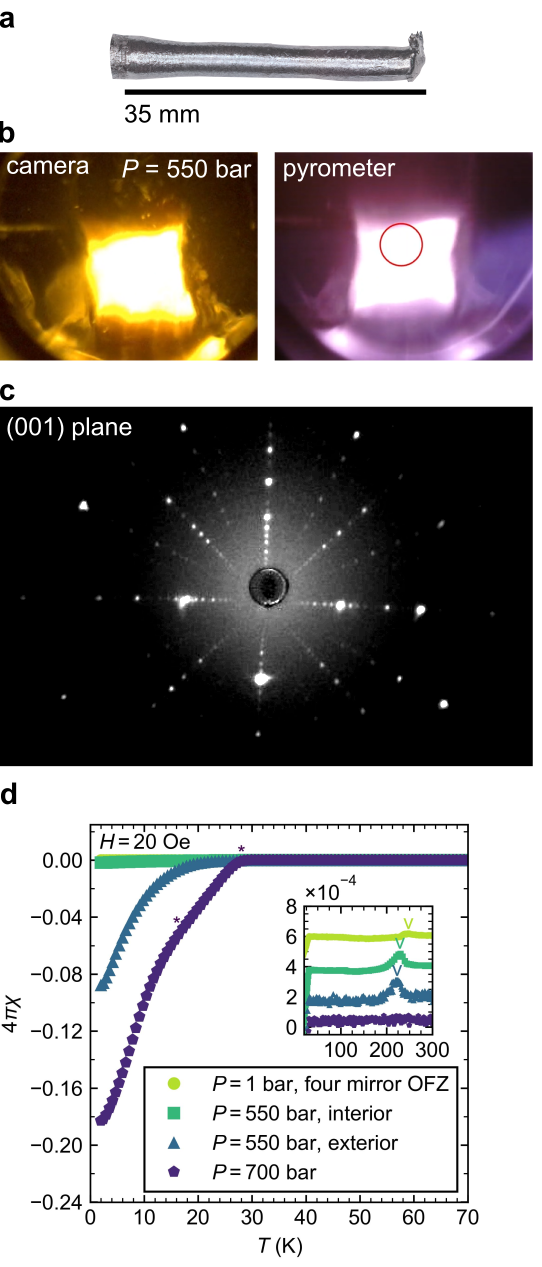}
		\caption{\label{fig:LCO}(a) Single crystal of La$_2$CuO$_{4+\delta}$ obtained from the first traveling solvent floating zone carried out in the HP-LFZ --- a shiny grey boule with an a subtle facet corresponding to the (001) plane. (b) Images of the growth from the camera (top) and pyrometer (bottom) at $P=550$~bar. (c) Laue diffraction pattern obtained from a diced and polished cross section of the boule, consistent with reflections from the (001) plane. (d) Temperature-dependent magnetic susceptibility data at $H=20$~Oe. Two superconducting transitions are marked with asterisks (*). An increase in superconducting volume fraction at higher pressures is attributed to enhanced O$_2$ intercalation. Inset shows variation in the Néel transition temperature, with the transitions marked by downward arrows.}
	\end{figure}
	
	Finally, we demonstrate the first traveling solvent floating zone growth at the high pressures achievable in the LOKII design by performing crystal growth of La$_2$CuO$_{4+\delta}$ at 700 bar.
	La$_2$CuO$_4$ is a canonical parent material to high-temperature superconductivity upon hole doping via cation substitution in La$_{2-x}$Sr$_x$CuO$_4$ and La$_{2-x}$Ba$_x$CuO$_4$, or via interstitial oxygen excess in La$_2$CuO$_{4+\delta}$. Superconductivity can be achieved at temperatures $T_\mathrm{c}\geq40~$K in La$_2$CuO$_{4+\delta}$.\cite{tarascon_superconductivity_1987} Because La$_2$CuO$_4$ decomposes peritectically near $1325~^{\circ}$C,\cite{oka_phase_1987} a flux must be employed to grow crystals from a melt. Thus, La$_2$CuO$_4$ is conventionally grown using the traveling solvent floating zone technique, in which a CuO-rich flux is used to dissolve the feed rod and precipitate out the grown crystal. Here, we achieve the first traveling solvent floating zone growth operated at 700 bar, and demonstrate an increased oxygen solubility in the solvent as a function of pressure.
	
	Tuning the doping level in La$_2$CuO$_{4+\delta}$ using oxygen excess is typically achieved by post-growth annealing in high-pressure O$_2$,\cite{rogers_identification_1988,hundley_phase_1990,mccarty_superconducting_1991,hirayama_superconducting_1998} electrochemical methods,\cite{grenier_new_1991,casan-pastor_evidence_2001} or chemical oxidation via an oxidizing agent.\cite{rudolf_semiconductorsuperconductor_1992,rial_room_1997,lan_structure_2000,takayama-muromachi_direct_1993} The oxygen intercalates between the CuO$_2$ layers and can undergo ordering phenomena in stages, much like the intercalation of chemical species between layers of graphite.\cite{wells_incommensurate_1997,lorenz_intrinsic_2002,liu_thermal_2005,mohottala_phase_2006,kusmartsev_transformation_2000,lee_neutron_2004,fratini_scale-free_2010} Traveling solvent floating zone growth of La$_2$CuO$_4$ requires some O$_2$ pressure (conventionally on the order of 1~bar) in order to stabilize the Cu$^{2+}$ oxidation state,\cite{revcolevschi_growth_1997,dabkowska_growth_2007,revcolevschi_crystal_1999} and this results in as-grown samples exhibiting a partial volume fraction of superconducting phase due to intercalation of excess oxygen into the lattice. 

	Here we employ a standard recipe, using a flux with molar composition of 4:1 CuO:La$_2$O$_3$ and a feed rod containing 2\% excess CuO to sustain the flux composition during growth. To prepare both the feed rod and the flux, we began by grinding and mixing stoichiometric amounts of La$_2$O$_3$ (99.99\%, Alfa Aesar) and CuO (99.9995\% Thermo Scientific) powders. This mixture was then loaded into a latex balloon and pressed into a pellet using 3275~bar in a cold isostatic press. 
	
	The feed rod precursors underwent a three-step firing process with intermittent grinding and pressing steps to avoid inhomogeneity. First, the feed pellet was fired at 800~$^{\circ}$C for 24~hours in air. After the first firing, the resulting material was then reground into a fine powder before pressing and firing again, this time at 900~$^{\circ}$C for another 24~hours in air. Finally, the pellet was reground,  pressed into a rod, and sintered at 1200~$^{\circ}$C for 10~hours under flowing O$_2$. In a similar fashion, the flux pellet was first fired at 800~$^{\circ}$C for 24~hours in air, then at 800~$^{\circ}$C for 24~hours in air, and lastly at 850~$^{\circ}$C for 12~hours in air.
	
	Prior to the growth, a piece of the flux was first fused to a single crystal seed rod in ambient atmosphere using a total melting power of $\approx$40~W, to avoid the challenge of balancing a disc of the flux on the seed while pressurizing the growth chamber. The growth was carried out under a total pressure of 700~bar of 80:20 Ar:O$_2$ mix, requiring a total power of 140~W to melt the flux. The rods were translated at a rate of 1~mm/hr, with rotation rates of 5 rpm and 10 rpm for the feed and seed, respectively. A stable zone was sustained for several days, producing multi-centimeter sized boules (Fig. \ref{fig:LCO}(a)). The surfaces of the crystals were shiny grey with a subtle facet visible on two opposite sides, and a Laue diffraction pattern on a piece diced from a boule in this orientation shows well-defined reflections corresponding to the (001) plane of the La$_2$CuO$_4$ phase (Fig. \ref{fig:LCO}(c)).
	
	Magnetization measurements on crystals diced from boules grown at pressures of 550 bar and 700 bar reveal two superconducting transitions near $T_{c1}=28$~K and $T_{c2}=16$~K (Fig. \ref{fig:LCO}(d)), suggestive of a mixed state where two different and unknown orderings of interstitial oxygen coexist in segregated regions of the sample.\cite{fratini_scale-free_2010} Compared to samples grown under 1~bar of O$_2$ in a conventional mirror furnace, crystals grown in the HP-LFZ exhibit a larger superconducting volume fraction, which we attribute to higher amounts of intercalated oxygen inherent to the growth process under higher pressures of O$_2$. To assess the distribution of oxygen within the boule, two pieces were measured for the 550~bar growth: one piece diced from the center of the boule and one piece diced from the exterior. The piece taken from the exterior exhibited a significantly higher volume fraction, suggesting a radial gradient in the doping level. Thermogravimetric analysis (TGA) measurements were performed on a crushed piece of the 700~bar growth in flowing N$_2$ to estimate the amount of intercalated oxygen ($\delta$) lost after heating as-grown La$_2$CuO$_{4+\delta}$ to 950~$^{\circ}$C. The resulting mass loss suggested an average composition $\delta=0.07$ of the as-grown boule.
	These results suggest that higher O$_2$ partial pressures lead to a increase in the activity of oxygen in the growth process compared to growth under conventional pressures.

	\section{Summary}
	A second generation design of the high-pressure laser floating zone furnace, LOKII, is presented.  This design advance uses an optics array that allows for in-plane imaging and higher power at the growth front. Additionally, the removal of out-of-plane sight glasses on the chamber allows for better cooling of the chamber and window housings. The growth of several representative oxides demonstrates the ability of a high-pressure growth environment to tune chemical potential and mitigate material volatility. We further demonstrate the feasibility of the traveling solvent floating zone technique at a pressure of 700 bar and show that wide band gap materials can be grown in regimes where initial coupling to the laser light is weak. This second-generation design is expected to aid in realizing the high quality single crystal growth of other volatile and metastable phases of interest in the field of quantum materials.
	
	\section{Acknowledgments}
	\noindent Jude Quirinale, \textit{in memoriam}.\newline
	
	The authors acknowledge fruitful conversations with Mitchell Bordelon, Zach Porter, Brenden R. Ortiz, Paul M. Sarte, Guang Wu, Bin Gao, Karthik Rao, Alon H. Avidor, Pedro A. Barrera, Jessica Jauregui-Gonzalez, and Casandra J. Gomez Alvarado. This work was supported via the UC Santa Barbara NSF Quantum Foundry funded via the Q-AMASE-i program under award DMR-1906325. S.J.G.A. acknowledges additional financial support from the National Science Foundation Graduate Research Fellowship under Grant No. 1650114. This research also made use of the shared facilities of the NSF Materials Research Science and Engineering Center at UC Santa Barbara, Grant No. DMR-1720256. 
	
	\section{Author Declarations}
 	\subsection{Conflict of Interest}
 	The authors have no conflicts to disclose.	
 	\subsection{Author Contributions}
 	\noindent\textbf{Steven J. Gomez Alvarado}: Investigation (lead); Methodology (lead); Software (lead); Validation (lead); Visualization (equal); Writing - original draft (lead). Writing - review \& editing (equal). 
 	\textbf{Eli Zoghlin}: Investigation (equal); Methodology (equal); Project administration (equal); Supervision (equal); Validation (equal); Writing - review \& editing (equal).
 	\textbf{Azzedin Jackson}: Methodology (equal); Validation (equal).
 	\textbf{Linus Kautzsch}: Investigation (equal); Methodology (equal); Validation (equal).
 	\textbf{Jayden Plumb}: Formal analysis (equal); Investigation (equal); Methodology (equal); Validation (equal); Visualization (equal).
 	\textbf{Michael Aling}: Formal analysis (lead); Investigation (equal); Methodology (equal); Visualization (lead).
 	\textbf{Andrea N. Capa Salinas}: Investigation (equal); Visualization (equal).
 	\textbf{Ganesh Pokharel}: Investigation (equal); Visualization (equal).
 	\textbf{Yiming Pang}: Investigation (equal); Methodology (equal); Validation (equal).
 	\textbf{Reina M. Gomez}: Investigation (equal); Methodology (equal); Validation (equal).
 	\textbf{Samantha Daly}: Supervision (equal); Writing - review \& editing (equal).
 	\textbf{Stephen D. Wilson}: Conceptualization (lead); Funding acquisition (lead); Project administration (lead); Resources (lead); Supervision (lead); Writing - review \& editing (equal).
 	
 	\section{Data Availability}
 	The data that support the findings of this study are available from the corresponding author upon reasonable request.

\end{document}